\documentclass[aps,floatfix,twocolumn,nofootinbib,showpacs]{revtex4}
\usepackage{graphicx}
\usepackage{latexsym}

\begin{document}

\title{Optimal Paths in Disordered Complex Networks}

\author{Lidia A. Braunstein,$^{1,2}$ Sergey V. Buldyrev,$^1$,
Reuven Cohen,$^3$ Shlomo Havlin,$^{1,3}$ and H. Eugene Stanley$^1$}

\affiliation{$^1$Center for Polymer Studies and Department of Physics, 
Boston University, Boston, MA 02215, USA\\
$^2$Departamento de F\'{\i}sica, Facultad de Ciencias Exactas y
Naturales, Universidad Nacional de Mar del Plata, Funes 3350,
$7600$ Mar del Plata, Argentina\\
$^3$Minerva Center and Department of Physics,
Bar-Ilan University, 52900 Ramat-Gan, Israel}
\pacs{89.75.Hc}
\date{bbchs.tex ~~~ LE 9030, submitted 2 May 2003, revised 19 August 2003}

\begin{abstract}

We study the optimal distance in networks, $\ell_{\mbox{\scriptsize
opt}}$, defined as the length of the path minimizing the total weight,
in the presence of disorder.  Disorder is introduced by assigning
random weights to the links or nodes.  For strong disorder, where the
maximal weight along the path dominates the sum, we find that
$\ell_{\mbox{\scriptsize opt}}\sim N^{1/3}$ in both Erd\H{o}s-R\'enyi
(ER) and Watts-Strogatz (WS) networks. For scale free (SF) networks,
with degree distribution $P(k) \sim k^{-\lambda}$, we find that
$\ell_{\mbox{\scriptsize opt}}$ scales as $N^{(\lambda - 3)/(\lambda -
1)}$ for $3<\lambda<4$ and as $N^{1/3}$ for $\lambda\geq 4$. Thus, for
these networks, the small-world nature is destroyed. For $2 < \lambda
< 3$, our numerical results suggest that $\ell_{\mbox{\scriptsize
opt}}$ scales as $\ln^{\lambda-1}N$. We also find numerically that for
weak disorder $\ell_{\mbox{\scriptsize opt}}\sim\ln N$ for both the ER
and WS models as well as for SF networks.

\end{abstract}

%\pacs{XX}

\maketitle

Recently much attention has been focused on the topic of complex
networks which characterize many biological, social, and communication
systems \cite{Albert02,DM02,PastorXX}. The networks can be visualized
by nodes representing individuals, organizations, or computers and by
links between them representing their interactions.

The classical model for random networks is the Erd\H{o}s-R\'enyi (ER)
model \cite{ER59,ER60}.  An important quantity characterizing networks
is the average distance (minimal hopping) $\ell_{\mbox{\scriptsize
min}}$ between two nodes in the network of total $N$ nodes.  For the
Erd\H{o}s-R\'enyi network, and the related, more realistic
Watts-Strogatz (WS) network \cite{WattsXX} $\ell_{\mbox{\scriptsize
min}}$ scales as $\ln N$ \cite{Bollobas}, which leads to the concept
of ``six degrees of separation''.

In most studies, all links in the network are regarded as identical
and thus the relevant parameter for information flow including
efficient routing, searching, and transport is
$\ell_{\mbox{\scriptsize min}}$. In practice, however, the weights
(e.g., the quality or cost) of links are usually not equal, and thus
the length of the optimal path minimizing the sum of weights is
usually longer than the distance. In many cases, the selection of the
path is controlled by the sum of weights (e.g., total cost) and this
case corresponds to regular or weak disorder. However, in other cases,
for example, when a transmission at a constant high rate is needed
(e.g., in broadcasting video records over the Internet) the narrowest
band link in the path between the transmitter and receiver controls
the rate of transmission. This situation---in which one link controls
the selection of the path---is called the strong disorder limit.  In
this Letter we show that disorder or inhomogeneity in the weight of
links may increase the distance dramatically, destroying the
``small-world'' nature of the networks.

To implement the disorder, we assign a weight or ``cost'' to each link
or node. For example, the weight could be the time $\tau_i$ required
to transit the link $i$.  The optimal path connecting nodes A and B is
the one for which $\sum_i\tau_i$ is a minimum. While in weak disorder
all links contribute to the sum, in strong disorder one term dominates
it. The strong disorder limit may be naturally realized in
the vicinity of the absolute zero temperature if passing through a
link is an activation process with a random activation energy
$\epsilon_i$ and $\tau_i=\exp(\beta\epsilon_i)$ , where $\beta$ is
inverse temperature\cite{disorder}.  Let us assume that the energy spectrum is
discrete and that the minimal difference between energy levels is
$\Delta\epsilon$. It can be easily shown that if $\beta>\ln
2/\Delta\epsilon$, the value of $\sum_i\tau_i$ is dominated by the
largest term, $\tau_{\mbox{\scriptsize max}}$. Thus if we have two
different paths characterized by the sums $\sum_i\tau_i$ and
$\sum_i\tau'_i$, such that $\tau_{\mbox{\scriptsize
max}}>\tau'_{\mbox{\scriptsize max}}$, it follows that
$\sum_i\tau_i>\sum_i\tau'_i$.

To generate ER graphs, we start with $zN$ links and for each link
randomly select from the total $N(N-1)/2$ possible pairs of nodes a
pair that is connected by this link. The WS network
\cite{WattsXX} is implemented by placing the $N$ nodes on a
circle. Initially, each node $i$ is connected with $z$ nodes
$i+1,i+2,...,i+z$ and periodic boundaries are implemented. Thus each
node has a degree $2z$ and the total number of links is $zN$.  Next we
randomly remove a fraction $p$ of the links and use them to connect
randomly-selected pairs of nodes.  When $p=1$, we obtain a model very
similar to the ER graph.

To generate scale-free (SF) graphs, we employ the Molloy-Reed algorithm
\cite{Molloy} in which each node is first assigned a random integer $k$
from a power law distribution $P(k>\bar k)=(\bar k/k_0)^{-\lambda+1}$,
where $k_0$ is the minimal number of links for each node. Next we
randomly select a node and try to connect each of its $k$ links with
randomly selected $k$ nodes that still have free positions for links. 
 
We expect that the optimal path length in the weak disorder case will
not be considerably different from the shortest path, as found for
regular lattices \cite{RednerXX} and random graphs
\cite{verderHofstad01}. Thus we expect that the scaling for the shortest
path $\ell_{\mbox{\scriptsize min}}\sim\ln N$ will also be valid for the
optimal path in weak disorder, but with a different prefactor depending
on the details of the graph.

In the case of strong disorder, we present the following theoretical
arguments. Cieplak et al.\cite{Cieplak} showed that finding the
optimal path between nodes A and B in the strong disorder limit is
equivalent to the following procedure.
First, we sort all $M$ links of the network in the descending order of
their weights, so that the first link in this list has the largest
weight.  Since the sum of the weights on any path between nodes A and
B is dominated by a single link with the largest weight, the optimal
path cannot go through the first link in the list, provided there is
a path between A and B which avoids this link. Thus the first
link in the list can be eliminated and now our problem is reduced to
the problem of finding the minimal path on the network of $M-1$
links. We can continue to remove links from the top of the list one by
one until we pick a link whose removal destroys the connectivity
between A and B.  This means that all the remaining paths between A
and B go through this singly-connecting or ``red'' link
\cite{ConiglioXX} and all these paths have the same largest weight
corresponding to the ``red'' link. To continue optimization among
these paths we must select the paths with the minimal second largest
term, minimal third largest term and so on. So we must continue to
remove links in the descending order of their weights unless they are
``red'' until a single path between A and B, consisting of only
``red'' links remains.  Since the assigning of weights to the links is
random so is their ordering.  Hence the optimization procedure in the
strong disorder limit is statistically equivalent to removing the
links in random order unless the connectivity between nodes A and B is
destroyed.

At the beginning of this
process, the chances of losing connectivity by removing a random link
are very low, so the process corresponds exactly to diluting the
network, which is identical to the percolation model. Only when the
concentration of the remaining links approaches the percolation
threshold will the chances of removing a singly-connected ``red'' link
\cite{ConiglioXX} become significant, indicating that the optimal path
must be on the percolation backbone connecting A and B. Since the
network is not embedded in space but has an infinite dimensionality,
we expect from percolation theory that loops are not relevant at
criticality~\cite{CEBH00}. 
Thus, the shortest path must also be the optimal path.

We begin by considering the case of the ER graph that, at criticality,
is equivalent to percolation on the Cayley tree or percolation at the
upper critical dimension $d_c=6$. For the ER graph, it is known that the
mass of the incipient infinite cluster $S$ scales as $N^{2/3}$
\cite{ER59}. This result can also be obtained in the framework of
percolation theory for $d_c=6$. Since $S\sim R^{d_f}$ and $N\sim R^d$
(where $d_f$ is the fractal dimension and $R$ the diameter of the
cluster), it follows that $S\sim N^{d_f/d}$ and for $d_c=6$, $d_f=4$
\cite{BH96}
\begin{equation}
S\sim N^{2/3}.
\label{E.1x}
\end{equation}
It is also known \cite{BH96} that, at criticality, at the upper critical
dimension, $S\sim\ell_{\mbox{\scriptsize min}}^{d_\ell}$ with $d_\ell=2$,
\cite{BH96}
and thus
\begin{equation}
\ell_{\mbox{\scriptsize min}} \sim \ell_{\mbox{\scriptsize opt}}\sim
S^{1/d_\ell}\sim N^{2/3d_\ell} \sim
N^{\nu_{\mbox{\scriptsize opt}}},
\label{E.1}
\end{equation}
where $\nu_{\mbox{\scriptsize opt}}=2/3d_\ell=1/3$.  We expect that the
WS model for large $N$ and large $p$ will be in the same universality
class as ER.

For SF networks, we can also use the percolation results at
criticality. It was found \cite{Cohen} that $d_\ell=2$ for $\lambda>4$,
$d_\ell=(\lambda-2)/(\lambda-3)$ for $3<\lambda<4$, $S\sim N^{2/3}$ for
$\lambda>4$, and $S\sim N^{(\lambda-2)/(\lambda-1)}$ for
$3<\lambda\leq 4$. Hence, we conclude that
\begin{equation}
\ell_{\mbox{\scriptsize min}}\sim
\ell_{\mbox{\scriptsize opt}}\sim\cases{
                    N^{1/3} & $\lambda>4$ \cr
N^{(\lambda-3)/(\lambda-1)} & $3<\lambda\leq 4$}.
\label{E.2x}
\end{equation}
To test these theoretical predictions, we perform numerical
simulations in the strong disorder limit by randomly removing links
(or nodes) for ER, WS, and SF networks and use the Dijkstra algorithm
\cite{Cormen90} for the weak disorder case. We also perform additional
simulations for the case of strong disorder on ER networks using
direct optimization with Dijkstra algorithm of the sum of weights and
find results identical to the results obtained by randomly removing
links, see Fig.~2a.

\begin{figure}
\includegraphics[width=5.5cm,height=7cm,angle=270]{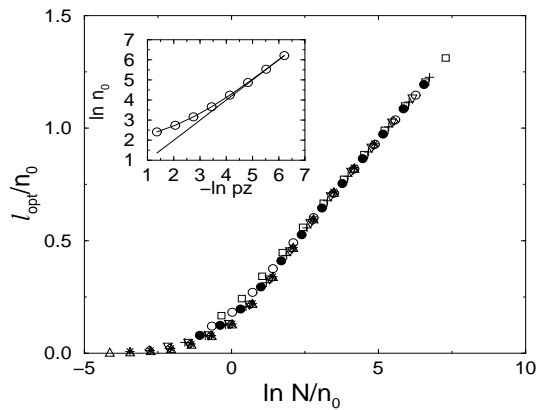}
\caption{Scaling plot of $\ell_{\mbox{\scriptsize opt}}$ on WS graphs
for weak disorder as a function $\ln N/n_0$ for various values of $p$ 
and $z=2$. The inset shows the log-log plot of $n_0$ versus $pz$.
The different symbols represent different $p$ values: 
$p=0.001$ ($\bigtriangleup$),
$p=0.002$ ($\ast$),
$p=0.004$ ($\diamond$),
$p=0.008$ ($\bigtriangledown$),
$p=0.016$ ($+$),
$p=0.032$ ($\bullet$)
$p=0.064$ ($\circ$), and
$p=0.128$ ($\Box$). Similar results have been obtained for $z=1$, 4, and
8. Those results scale according to Eq.~(\protect\ref{e1x}).
\label{f.1}}
\end{figure}

\begin{figure}
%\centerline{
\includegraphics[width=5.5cm,height=7cm,angle=270]{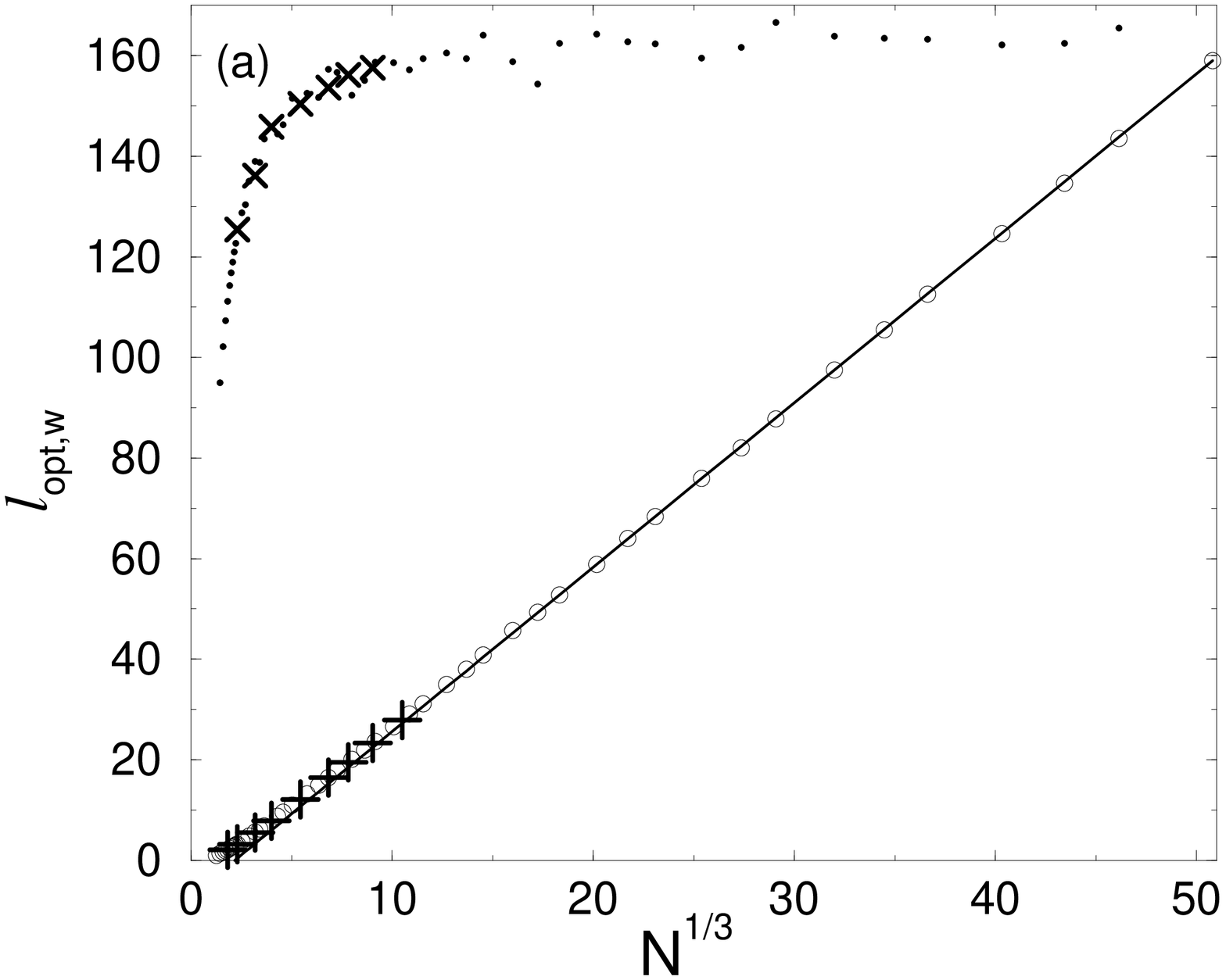}
\includegraphics[width=5.5cm,height=7cm,angle=270]{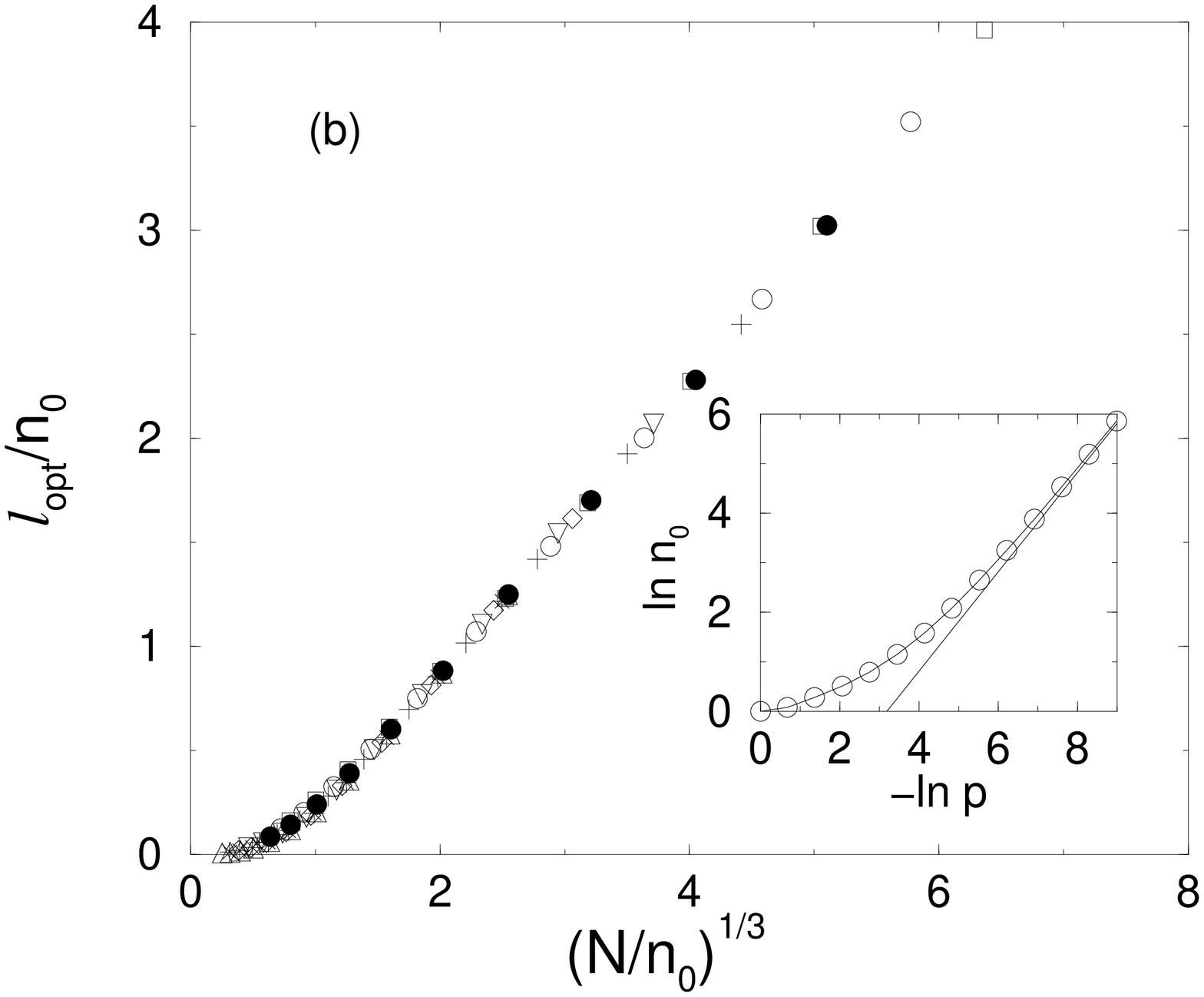}
%}
\caption{(a) The dependence of $\ell_{\mbox{\scriptsize opt}}$ on
$N^{1/3}$ for ER graphs for the strong disorder case obtained by
direct optimization ($+$) and by randomly removing links ($\circ$).
The linear asymptote has a slope of 3.27.  Also shown are the
successive slopes multiplied by 50 for direct optimization ($\times$)
and for randomly removing links ($\bullet$).  (b) Scaling plot of
$\ell_{\mbox{\scriptsize opt}}$ in WS graphs for strong disorder as a
function $(N/n_0)^{1/3}$ for various values of $p$ and $z=2$. The
symbols indicating values of $p$ are the same as in Fig. 1. The inset
shows a log-log plot of $n_0$ versus $p$ for $z=2$.
\label{f.2}}
\end{figure}

Results for weak disorder for WS graphs with different $p$ are shown
in Fig.~1. We propose a scaling formula for $\ell_{\mbox{\scriptsize
opt}}$ similar to the formula derived in \cite{Amaral99,Newman} for
the minimal distance on the WS graphs with a different rewiring
probability $p$
\begin{equation}
\label{e1x}
\ell_{\mbox{\scriptsize opt}} \sim 
{n_0(p,z)\over n_1(z)}F\left({N\over n_0(p,z)}\right),
\end{equation}
where $n_0(p,z) \sim 1/pz$ is the characteristic graph size at which the
crossover from large to small world behavior occurs, $n_1(z) \sim z$ is
a correction factor, and $F(x)$ is the scaling function
\begin{equation}
\label{e2x}
F(x)\sim\cases{\ln x & $x\to\infty$\cr
                   x  & $x\to 0$}.
\end{equation}
The scaling variable $x=N/n_0$ indicates the number of nodes with long
range links. As $p \rightarrow 0$, this quantity scales as $Npz$. The
quantity $n_0(p,z) \sim 1/pz$ indicates a typical short-range
neighborhood of a node with long range links. We can think about this
graph as an ER graph consisting of $N/n_0$ effective nodes, each
representing a typical short-range chain-like neighborhood of size
$n_0$. Thus we conclude that an optimal path connecting any two
nodes is proportional to $\ln(N/n_0)$, as in an ER graph, times an
average path length through a chain of short-range links. This average
path is proportional to the length of this chain $n_0$ and inversely
proportional to the average range $n_1(z)$ of a link in this
chain. Ideally $n_1(z)=z$, but in reality it can significantly deviate
from $z$ due to finite size effects.  Figure~1 shows the scaled
optimal path $\ell_{\mbox{\scriptsize opt}}/n_0$ versus the scaled
variable $N/n_0$ for $z=2$ and different values of $p$. The inset in
Fig.~1 shows that $n_0 \sim 1/pz$ as $p\rightarrow 0$.

\begin{figure}
%\centerline{
\includegraphics[width=3.5cm,height=4.0cm,angle=270]{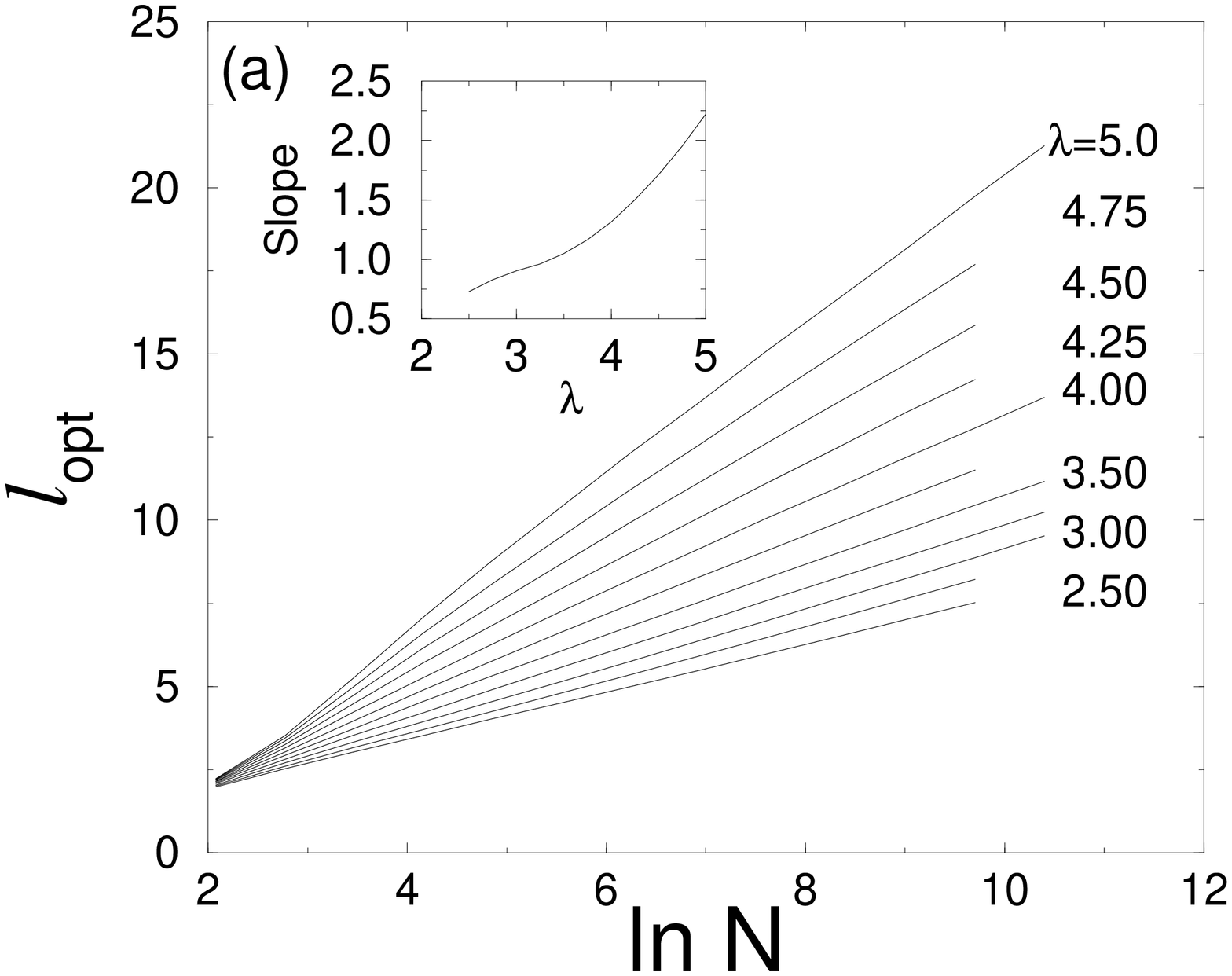}
\includegraphics[width=3.5cm,height=4.0cm,angle=270]{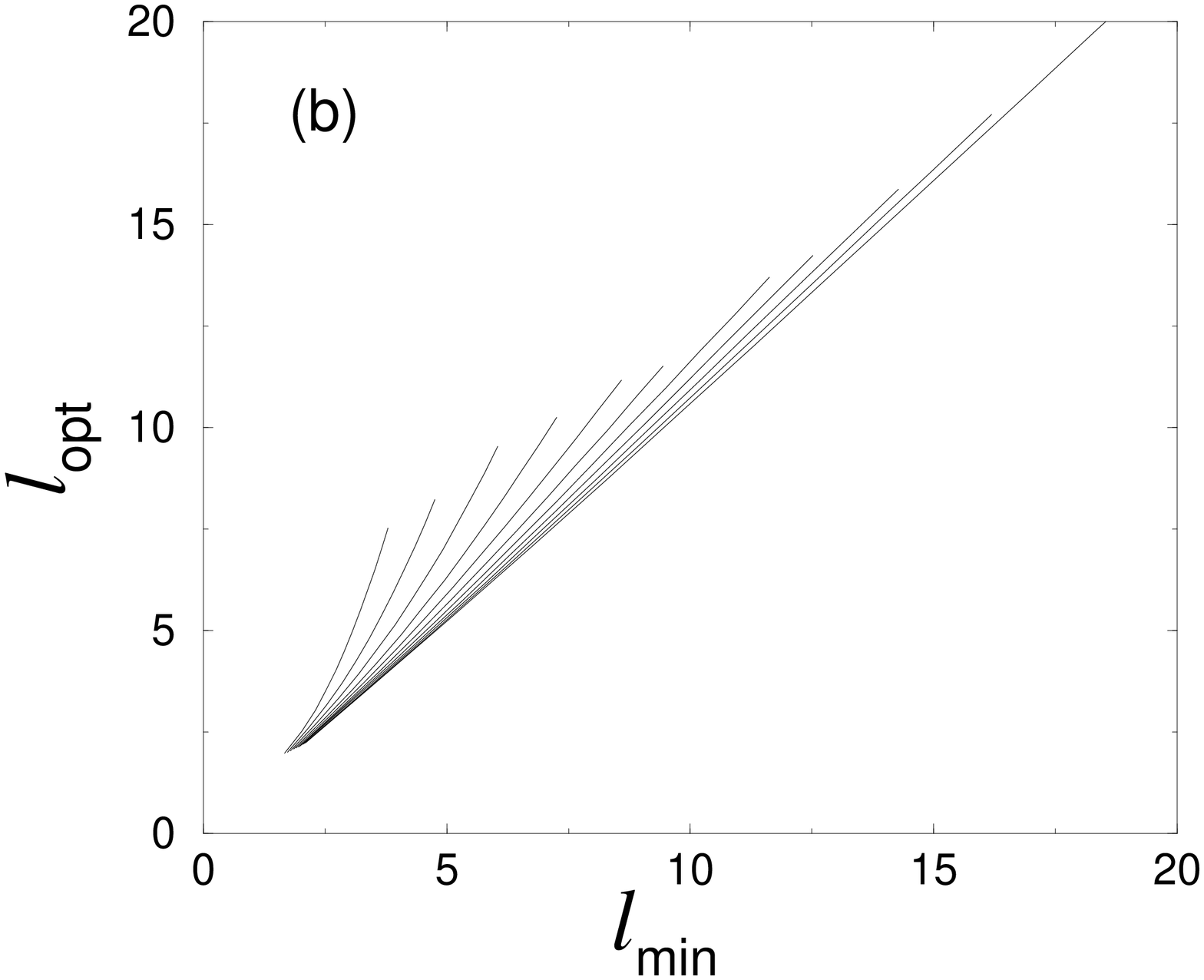}
%}
\caption{(a) The dependence of $\ell_{\mbox{\scriptsize opt}}$ on $\ln N$
for SF graphs in the weak disorder case for various values of $\lambda$
shown on the graph. The behavior of the asymptotic slope versus $\lambda$
shown as an inset. (b) The dependence of $\ell_{\mbox{\scriptsize opt}}$
on $\ell_{\mbox{\scriptsize min}}$. The curves from left to right represent
increasing values of $\lambda$ given in (a).
\label{f.3}}
\end{figure}

\begin{figure}
\includegraphics[width=3.5cm,height=4cm,angle=270]{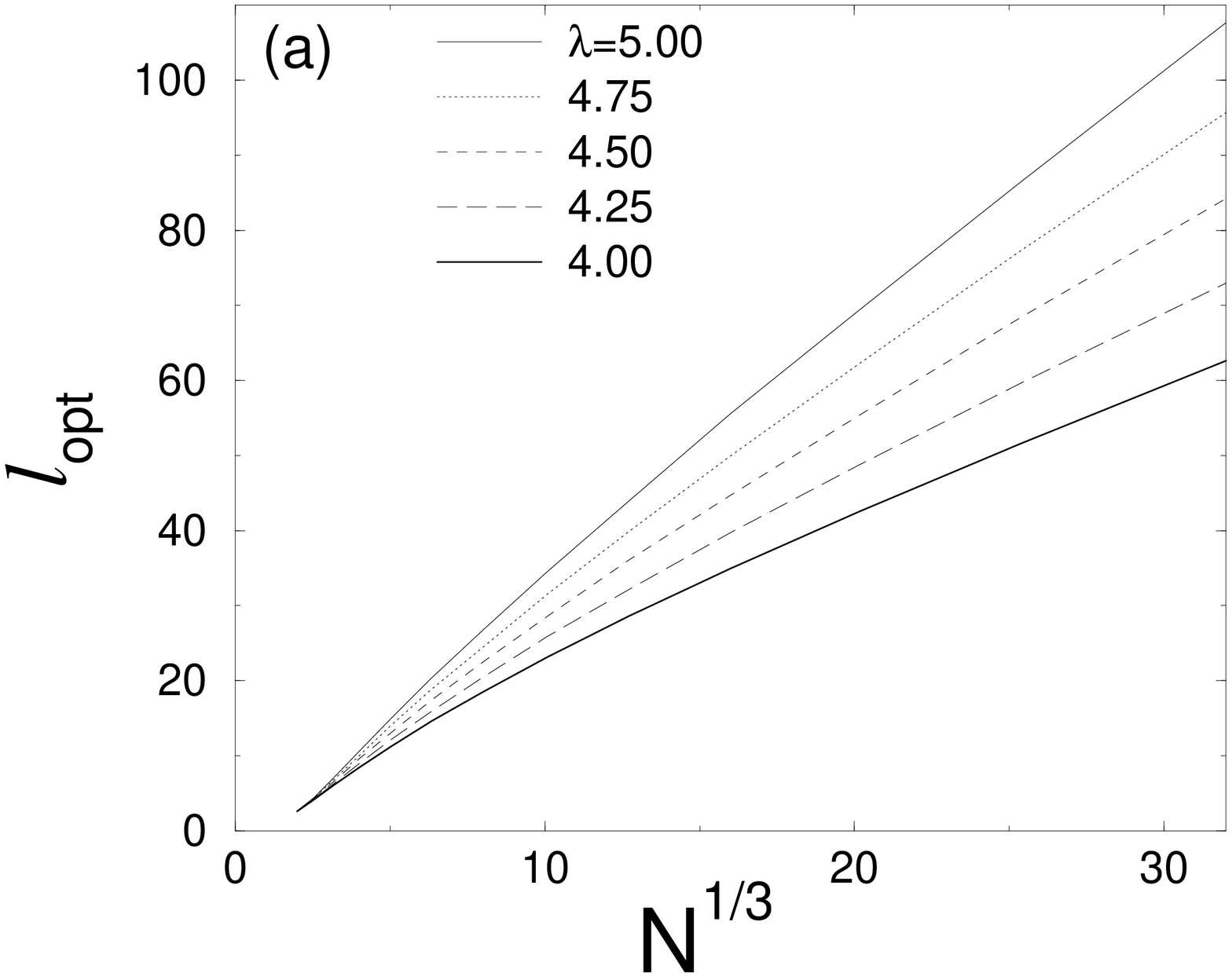}
\includegraphics[width=3.5cm,height=4cm,angle=270]{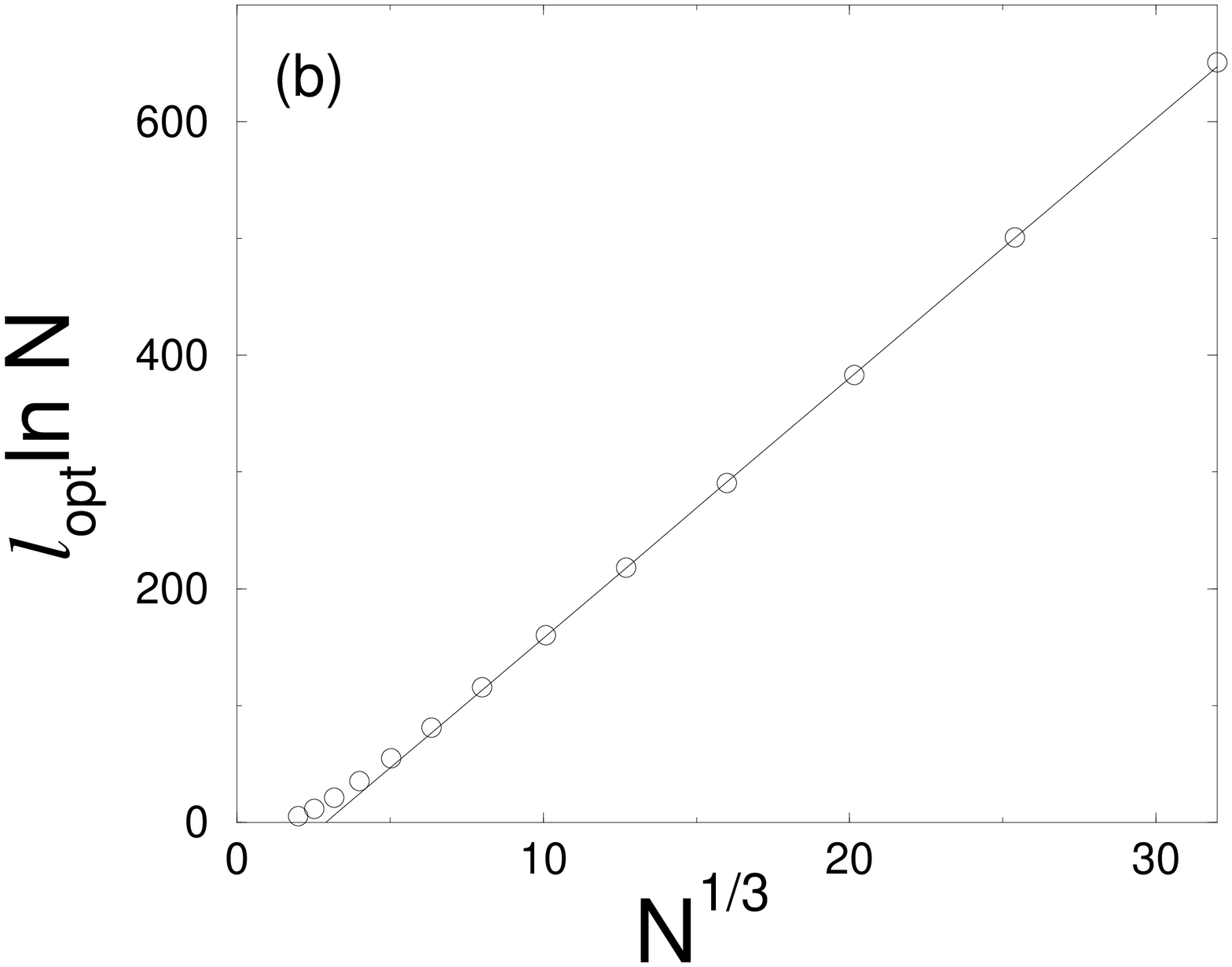}
\eject
\includegraphics[width=3.5cm,height=4cm,angle=270]{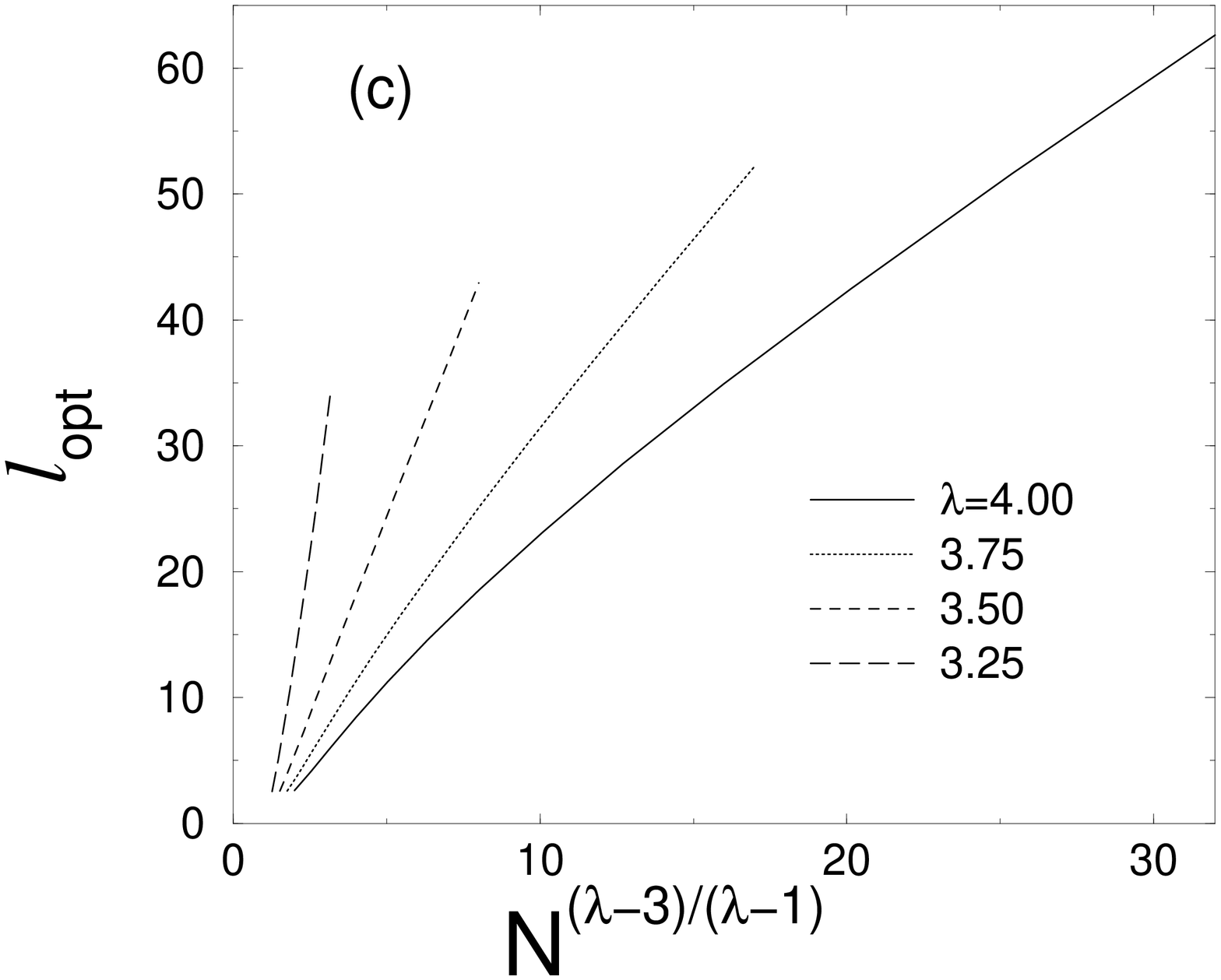}
\includegraphics[width=3.5cm,height=4cm,angle=270]{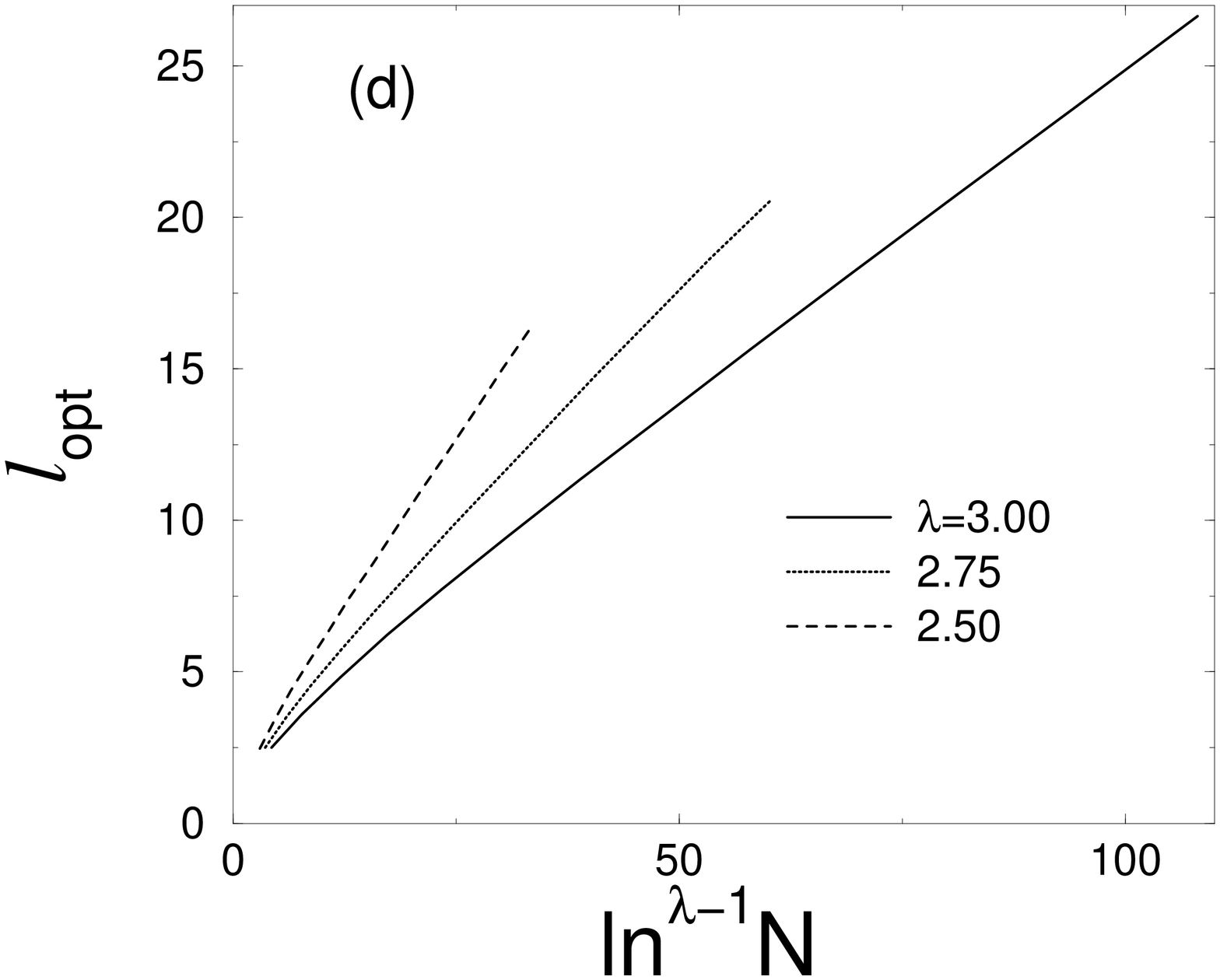}
\eject

\caption{
(a) The dependence of $\ell_{\mbox{\scriptsize opt}}$ on $N^{1/3}$ for
$\lambda\geq 4$.
(b) The dependence of $\ell_{\mbox{\scriptsize opt}}/\ln N$ on $N^{1/3}$ for 
$\lambda =4$. 
(c) The dependence of $\ell_{\mbox{\scriptsize opt}}$ on 
$N^{(\lambda-3)/(\lambda - 1)}$ for $3< \lambda <4$.
(d) The dependence of $\ell_{\mbox{\scriptsize opt}}$ on $\ln N$ for 
$\lambda\leq 3$. 
\label{f.4}}
\end{figure}

In contrast, for Eq.~(\ref{e1x}) to be in agreement with Eq.~(\ref{E.1})
for the strong disorder limit, we have (see Fig.~2),
\begin{equation}
\label{e3x}
F(x)\sim\cases{x^{1/3} & $x\to\infty$\cr
                   x  & $x\to 0$}.
\end{equation}
For large enough $z$ and $p\rightarrow 1$,  
we recover ER network for which $\ell_{\mbox{\scriptsize opt}}$ does not
depend on $z$. Thus we can assume $n_0(1,z)=1$. Using similar scaling arguments
as in case of weak disorder, we assume that as $p\rightarrow 0$,
$\ell_{\mbox{\scriptsize opt}}\sim z^{-2/3}N$, and hence $n_1(z)\sim z^{2/3}$.

For SF networks, the behavior of the optimal path in the weak
disorder limit is shown in Fig.~3 for different degree distribution
exponents $\lambda$. Here we plot $\ell_{\mbox{\scriptsize opt}}$ as a
function of $\ln N$. All the curves have linear asymptotes, but the
slopes depend on $\lambda$,
\begin{equation}
\label{e4x}
\ell_{\mbox{\scriptsize opt}}\sim f(\lambda)\ln N.
\end{equation}
This result is analogous to the behavior of the shortest path
$\ell_{\mbox{\scriptsize min}}\sim\ln N$ for $3<\lambda<4$. However,
for $2<\lambda<3$, $\ell_{\mbox{\scriptsize min}}$ scales as $\ln\ln
N$ \cite{Cohen3} while $\ell_{\mbox{\scriptsize opt}}$ is
significantly larger and scales as $\ln N$ (Fig.~3b). Thus weak
disorder does not change the universality class of the length of the
optimal path except in the case of ``ultra-small'' worlds
$2<\lambda<3$.

In contrast, strong disorder dramatically changes the universality class
of the optimal path. Theoretical considerations [Eqs.~(\ref{E.1}) and
(\ref{E.2x})] predict that in the case of WS and ER (Fig.~2) and
SF graphs with $\lambda>4$, $\ell_{\mbox{\scriptsize
opt}}=N^{1/3}$, while for SF graphs with $3<\lambda<4$,
$\ell_{\mbox{\scriptsize opt}}\sim N^{(\lambda - 3)/(\lambda -
1)}$. Figure~4a shows the linear behavior of $\ell_{\mbox{\scriptsize
opt}}$ versus $N^{1/3}$ for $\lambda\geq 4$.  The quality of the linear fit
becomes poor for $\lambda \rightarrow 4$. At this value, the logarithmic
divergence of the second moment of the degree distribution occurs and
one expects logarithmic corrections, i.e., $\ell_{\mbox{\scriptsize
opt}} \sim N^{1/3}/\ln N$ (see Fig.~4b).  Figure~4c shows the asymptotic
linear behavior of $\ell_{\mbox{\scriptsize opt}}$ versus $N^{(\lambda -
3)/(\lambda - 1)}$ for $3<\lambda\leq 4$.  Theoretically, as $\lambda
\rightarrow 3$, $\nu_{\mbox{\scriptsize opt}} =(\lambda - 3)/(\lambda
-1) \rightarrow 0$, and thus one can expect for $\lambda=3$ a
logarithmic dependence of $\ell_{\mbox{\scriptsize opt}}$ versus $N$.
Interestingly, for $2<\lambda<3$ our numerical results for the strong
disorder limit suggest that $\ell_{\mbox{\scriptsize opt}}$ scales
faster than $\ln N$. The numerical results can be fit to
$\ell_{\mbox{\scriptsize opt}}\sim(\ln N)^{\lambda-1}$ (see Fig.~4d). Note
that the correct asymptotic behavior may be different and this result
represents only a crossover regime. We obtain the same results for the
SF networks in which the weights are associated with nodes rather then
links. The exact nature of the percolation cluster at $\lambda<3$ is
not clear, since in this regime the tarnsition does not occur at a 
finite concentration~\cite{CEBH00}.

In summary, we study the optimal distance in ER, WS, and SF networks
in the presence of strong and weak disorder.  We find that in ER and
WS networks and for strong disorder, the optimal distance
$\ell_{\mbox{\scriptsize opt}}$ scales as $N^{1/3}$. We also study the
strong disorder limit in SF networks theoretically and by
simulations and find that $\ell_{\mbox{\scriptsize opt}}$ scales as
$N^{1/3}$ for $\lambda > 4$ and as $N^{(\lambda - 3)/(\lambda - 1)}$
for $3 < \lambda < 4$. Thus, the optimal distance increases dramatically in
strong disorder when it is compared to the known small world result
$\ell_{\mbox{\scriptsize min}}\sim\ln N$
and the ``small world'' nature for these networks is destroyed.  Our
simulations also suggest that for $2<\lambda<3$,
$\ell_{\mbox{\scriptsize opt}}$ scales as $\ln^{\lambda-1}N$, which
is also much faster than the ``ultra-small world'' result
$\ell_{\mbox{\scriptsize min}}\sim\ln (\ln N)$ \cite{Cohen3}.  We also
find numerically that in weak disorder $\ell_{\mbox{\scriptsize
opt}}\sim\ln N$ in all types of networks studied. 

\subsubsection*{Acknowledgments}

We thank A.-L. Barab\'asi and S. Sreenivasan for helpful discussions,
and ONR and Israel Science Foundation for financial support.


\begin{thebibliography}{99}
\bibitem{disorder}
We simulate weak disorder by selecting
$0\leq\tau_i<1$ from a uniform distribution. We simulate strong
disorder by selecting $0\leq\epsilon_i<1$ from a uniform distribution,
using a 48-bit random number generator, so that there are no two
identical values of $\epsilon_i$ in a system of any size that we
study. In this case $\Delta\epsilon\geq 2^{-48}$ and we can select
$\beta\geq 2^{48}$ to guarantee the strong disorder limit.
\end{thebibliography}
\end{document}